\documentclass[aps,twocolumn,showpacs,preprintnumbers,amsmath,amssymb]{revtex4}
\usepackage{graphicx}

\begin{document}

\title{Transfer of an unknown quantum state, quantum networks, and memory}
\author{Asoka Biswas and G. S. Agarwal}
\affiliation{Physical Research Laboratory, Navrangpura, Ahmedabad - 380 009, India}
\date{\today}

\begin{abstract}
We present a protocol for transfer  of an unknown quantum state.
The protocol is based on  a two-mode cavity 
interacting dispersively in a sequential manner with three-level atoms in 
$\Lambda$-configuration. 
We propose a scheme for quantum networking using an atomic channel.
We investigate the effect of cavity decoherence in the entire process.
Further we demonstrate the possibility of an efficient quantum memory for 
arbitrary superposition of two modes of a cavity containing one photon.

\end{abstract}

\pacs{03.67.-a, 03.67.Hk}

\maketitle

\section{Introduction}

In the quantum information theory \cite{info}, transfer of information in
the form of a coherently prepared quantum state is essential.
One can transfer a quantum state either by the method of teleportation 
\cite{bennett} or through quantum networking. 
The basic idea behind quantum network is to transfer a quantum state from one 
node to another node with the help of a career (a quantum channel) such that it
arrives intact. In between, one has to perform a process of quantum state transfer
(QST) to transfer the state from one node to the career and again from the career 
to the destination node. 
There have been some proposals \cite{cirac} for quantum networking 
using cavity-QED, where two atoms trapped inside two spatially separated 
cavities serve the purpose of two nodes. In Ref.~\cite{cirac} the task was to transfer the state
of {\it one atom into the other} via the process of QST
between the atom and photon where the later is used as a career. The
photon carries the information through either free space or an optical fiber
between the cavities and the success depends on the {\it probabilistic\/}
detection of photons or adiabatic passage through the cavities.
We note that, though it may be difficult to beat the communication with photons,
it is always interesting to explore the alternatives. In fact, very
recently, quantum network using linear XY chain of $N$ 
interacting qubits has been
proposed. In this proposal, the quantum state can be transferred from first qubit to $N$th
qubit within microscopic distance by pre-engineering interqubit interactions 
\cite{ekert}.

Further, storage of quantum states is also an 
important issue. There have been several proposals for quantum memory. 
For example, recent proposals \cite{lukin,molmer} have shown how to transfer the
field state into atomic coherence by adiabatic technique  and again retrieve the
same through the method of adiabatic following \cite{lukin} or using 
teleportation technique \cite{molmer}. Quantum memory of individual 
polarization state into a collective atomic ensemble has been proposed \cite{guo}.
Initially an entangled state of two pairs of atomic ensembles is prepared, where
the single photon polarization state is stored through a process similar to 
teleportation. Though the information can be transferred back to the photon state,
the protocol only succeeds with a probability 1/4.
Decoherence-free memory of one qubit in a pair of trapped ions has also been experimentally
demonstrated \cite{memo}. Ma\^{\i}tre {\it et al.\/} \cite{memory} have proposed a quantum 
memory, where the quantum information on the superposition state of
a two-level atom was stored in a 
cavity as a superposition of $0$ and $1$ photon Fock states. The
holding time of such memories is generally limited by the cavity decay time. 

In this paper, we propose a new scheme for QST to transfer the unknown state 
of one atom to another atom where the atoms are not 
directly interacting with each other. Note that by direct spin interaction
of $\vec{S}_1.\vec{S}_2$ kind, quantum 
state could be transferred from one atom to another within a microscopic range.
In the present scheme we show how similar kind of interaction between two atoms 
can be mediated via a cavity. Thus the atomic state can be transferred
from one atom to another in mesoscopic range. 

We extend our idea of QST to quantum network where
we transfer the {\it state of one cavity 
to another spatially separated cavity}. For this we use long-lived atoms as career,
and make use of the QST process to transfer the state of cavity to atom and again to the
target cavity. Our protocol
for quantum networking provides a {\it deterministic\/} way to transfer
the quantum state between the
cavities. This protocol does not require any kind of probability arguments
based on the outcome of a measurement.
Further we propose the realization of a quantum memory of {\it arbitrary superposition of two modes\/} of 
a cavity which contains only one photon. This
superposition state can be stored in the 
long-lived states of the neutral
atoms and retrieved in another two-mode cavity later, {\it deterministically}.
Our proposal relies on the technological advances and realizations as described in Ref.~\cite{haroche}. 

The structure of the paper is as follows. In Sec.~II, we describe the model and provide the 
relevant equations. In Sec.~III, we discuss how transfer of an
unknown quantum state can be performed between two atoms. We provide an estimate of possible
decoherence in this process due to cavity decay. In Sec.~IV, we extend our scheme to quantum
networks and quantum memory.

\section{Model configuration}

\begin{figure}
\scalebox{0.5}{\includegraphics{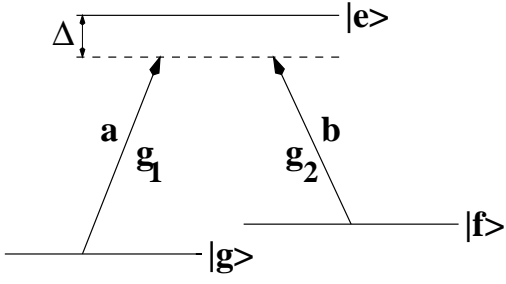}}
\caption{\label{fig1}Three-level atomic configuration with levels $|g\rangle$, 
$|e\rangle$, and $|f\rangle$ interacting with two orthogonal modes of the 
cavity, described by operators $a$ and $b$. Here $g_1$ and $g_2$ represent the 
atom-cavity coupling of the $a$ and $b$ modes with the corresponding 
transitions and $\Delta$ is the common one-photon detuning.}
\end{figure}

To describe how the QST protocol works,  we consider a three-level atom
in $\Lambda$ configuration interacting with a two-mode cavity (see Fig.~\ref{fig1}). 
The modes with annihilation operators $a$ and $b$ interact with the 
$|e\rangle \leftrightarrow |g\rangle$ and $|e\rangle \leftrightarrow |f\rangle$ 
transitions, respectively.
 The Hamiltonian under rotating wave approximation can be 
written as
\begin{eqnarray}
H&=&\hbar\left[\omega_{eg}|e\rangle\langle e|+\omega_{fg}|f\rangle\langle f|+\omega_1a^\dag a+\omega_2b^\dag b\right.\nonumber\\
&&+\left.\left\{g_1|e\rangle\langle g|a+g_2|e\rangle\langle f|b+\mathrm{H.c.}\right\}\right]
\label{fullhamil}
\end{eqnarray}
where $\omega_{lg}$ $(l \in e,f)$ is the atomic transition frequency, $\omega_i$
$(i\in 1,2)$ is the frequency of the cavity modes $a$ and $b$, and 
$g_i$ is the atom-cavity coupling constant. We assume $g_i$ to be real. 
 
We work under the two-photon resonance condition and assume large single-photon
detuning.
After adiabatically eliminating the excited level $|e\rangle$ in large detuning 
domain, we derive an effective Hamiltonian describing the system of Fig.~\ref{fig1}
\begin{eqnarray}
H_{\mathrm{eff}}&=&-\frac{\hbar g^2}{\Delta}\left[|g\rangle\langle g|a^\dag a+|f\rangle\langle f|b^\dag b\right]\nonumber\\
&&-\frac{\hbar g^2}{\Delta}\left[|g\rangle\langle f|a^\dag b+|f\rangle\langle g|a b^\dag\right],
\label{effham}
\end{eqnarray}
where $\Delta=\omega_{eg,f}-\omega_{1,2}$ is the common one-photon detuning 
of the cavity modes and $g_1=g_2=g$ $(\ll \Delta)$.
The condition $g_1=g_2$ can be satisfied by proper choice as we can choose
appropriate transitions in atomic systems, frequencies etc. 
Note that if one consider the levels $|g\rangle$ and 
$|f\rangle$ as Zeeman sublevels, then these conditions are automatically 
satisfied. In that case we may consider the two modes of the cavity as two 
orthogonal polarization states of a photon. 
Now note that, the first two terms in Eq.~(\ref{effham}) represent the self-energy terms and the last two terms
give the interaction leading to a transition from the initial state to the 
final state. 
The probability amplitudes of relevant basis states $|g\rangle |n,\mu\rangle$ 
and $|f\rangle|n-1,\mu +1\rangle$ in the state vector

\begin{equation}
|\psi(t)\rangle=d_g(t)|g,n,\mu\rangle+d_f(t)|f,n-1,\mu +1\rangle
\end{equation}
are given by

\begin{eqnarray}
d_g(t)&=&\frac{\sqrt{n}XY}{n+\mu+1}+d_g(0)\;,\nonumber\\
d_f(t)&=&\frac{\sqrt{\mu+1}XY}{n+\mu+1}+d_f(0)\;,
\label{solns}
\end{eqnarray}
where $X=\sqrt{n}d_g(0)+\sqrt{\mu+1}d_f(0)$, $Y=\exp[ig^2(n+\mu+1)t/\Delta]-1$,
 $n$ and $\mu$ are the respective photon numbers in the modes $a$ and $b$.
We note that the effective interaction (\ref{effham}) can be seen as an interaction
between two qubits defined via the atomic variables and field variables
\begin{eqnarray}
&&S^+=|f\rangle\langle g|,\;S^-=|g\rangle\langle f|,\;S^z=\frac{1}{2}(|f\rangle\langle f|-|g\rangle\langle g|);\nonumber\\
&&R^+=a^\dag b,\;R^-=ab^\dag,\;R^z=\frac{1}{2}(a^\dag a-b^\dag b).
\end{eqnarray}
In the single photon space, the field operators $R^\pm$, $R^z$ 
satisfy spin-$1/2$ algebra and thus the interaction (\ref{effham}) can be written as
interaction between two qubits
\begin{equation}
H_{\mathrm{eff}}\equiv -\frac{\hbar g^2}{\Delta}(R^+S^-+R^-S^+-2R^zS^z).
\end{equation}
In view of the above form of the effective interaction we conclude that our 
system of Fig.~\ref{fig1} can be used for a {\it number of quantum logic operations\/}.

\begin{figure}
\scalebox{0.32}{\includegraphics{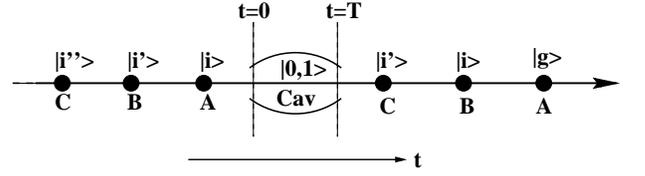}}
\caption{\label{fig2} Schematic diagram for the QST protocol
for a number of atoms interacting with the two-mode cavity  in a sequential manner
for a time $T=\Delta\pi/2g^2$.}
\end{figure}

\section{Quantum state transfer protocol}
We next demonstrate how the dynamics of an atom in a two-mode cavity can be 
used to implement the QST protocol. Now onwards we refer 
a $\pi$ pulse to an 
equivalent traversal time $T$ of the atom through the cavity such that, 
$2g^2T/\Delta =\pi$. The time $T$ could be controlled by selecting the atomic
velocity.

We assume that the atom A  is initially in an unknown state 
\begin{equation}
|i\rangle_A=\alpha |g\rangle_A +\beta |f\rangle_A\;,
\end{equation}
where $\alpha$ and $\beta$ are {\it unknown\/} arbitrary coefficients. The state $|i\rangle_A$ 
of atom A is to be transferred to another atom B  which is 
elsewhere. Preparing the cavity in a state $|0,1\rangle$ (i.e., initially one photon 
in the $b$ mode), we send the atom A through 
the cavity for certain time which is equivalent to a $\pi$ pulse. After atom A
comes out of the cavity, the atom B in state 
\begin{equation}
\label{state2}|i'\rangle=\alpha'|g\rangle+\beta'|f\rangle
\end{equation}
is sent through the cavity. Here $\alpha'$ and $\beta'$ are arbitrary 
coefficients and need not to be known. The atom B also experiences 
a $\pi$ pulse 
during the interaction with the cavity. The entire process can be described 
as follows:
\begin{eqnarray}
&|i\rangle_A&|0,1\rangle\nonumber\\
&\downarrow&~~ \pi~\textrm{pulse on atom A}\nonumber \\
&|g\rangle_A&(\alpha |0,1\rangle -\beta |1,0\rangle)\nonumber\\
\label{qst}&\downarrow&~~ \textrm{B atom enters}\\
&|g\rangle_A&|i'\rangle_B(\alpha |0,1\rangle -\beta |1,0\rangle)\nonumber\\
&\downarrow& ~~\pi~\textrm{pulse on atom B} \nonumber\\
&|g\rangle_A&|i\rangle_B(\alpha'|0,1\rangle -\beta'|1,0\rangle)\;.\nonumber
\end{eqnarray}

If one prepares the cavity initially in state $|1,0\rangle$, then following the 
similar sequence as above, the final state will be $-|f\rangle_A|i\rangle_B(\alpha'|0,1\rangle -\beta'|1,0\rangle)$.
Note that the atom B has already acquired the state $|i\rangle$ of atom A, i.e., the 
state $|i\rangle$ is transferred from the atom A  to atom B. 

\begin{figure*}
\begin{tabular}{cc}
\scalebox{0.35}{\includegraphics{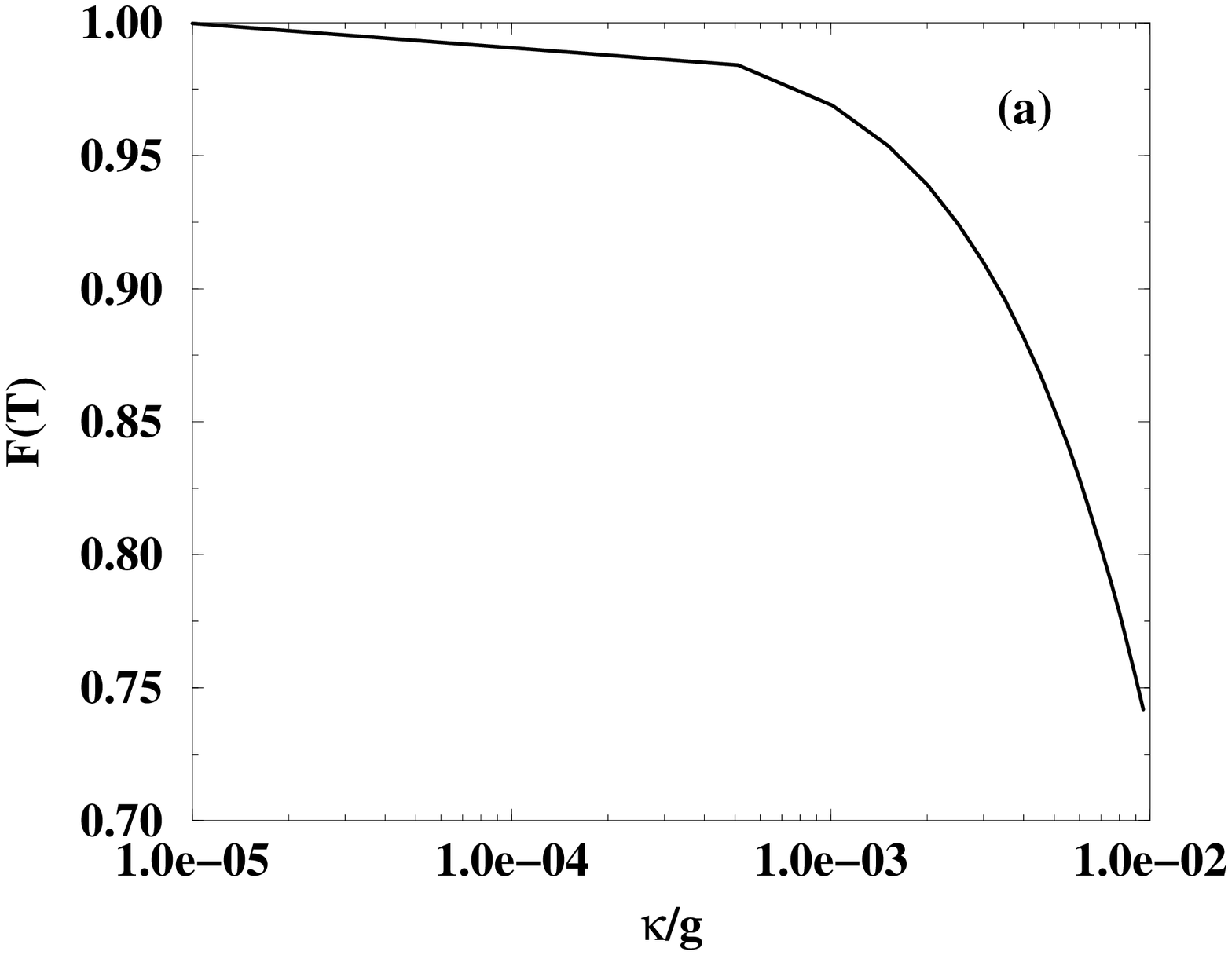}}&
\scalebox{0.35}{\includegraphics{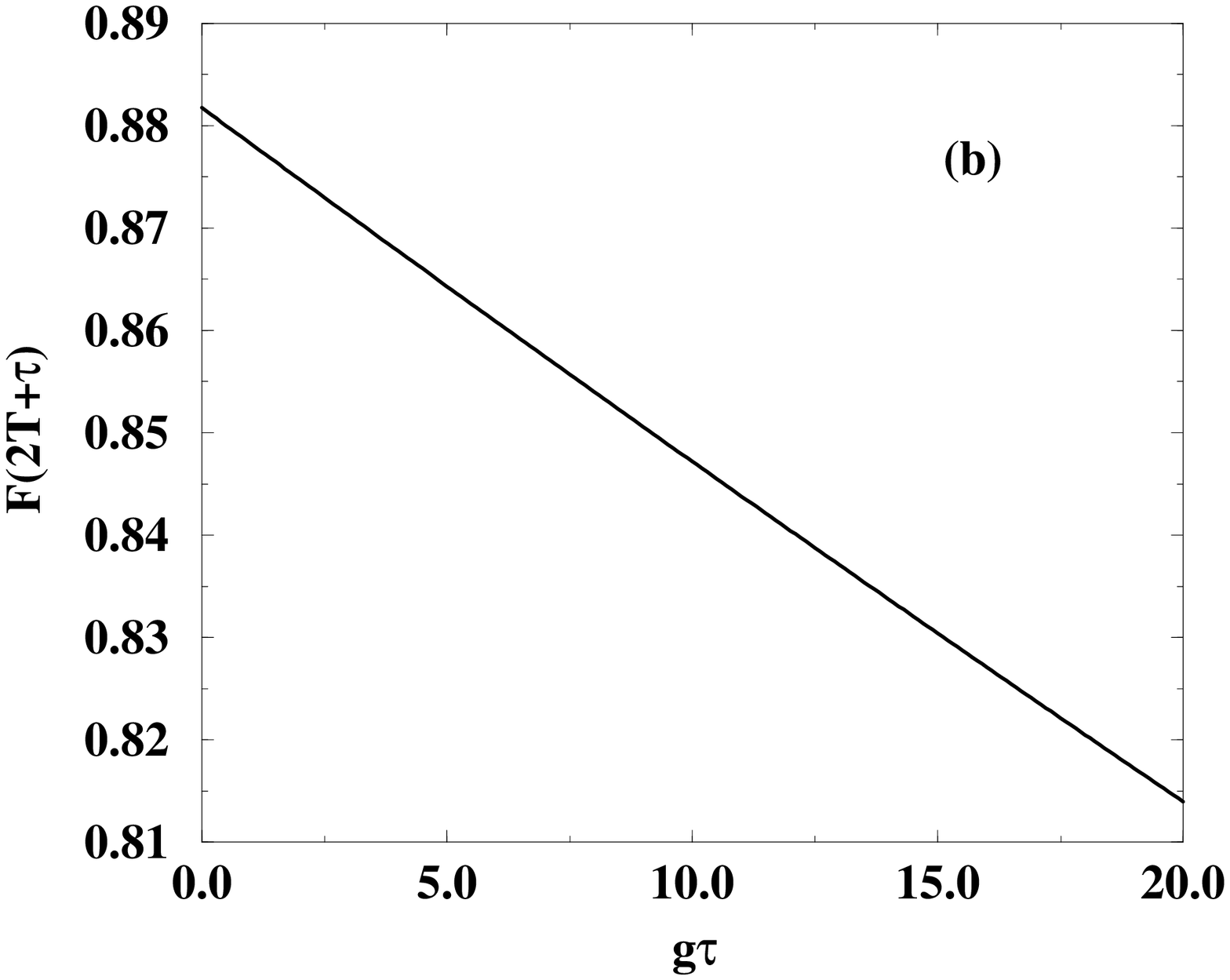}}
\end{tabular}
\caption{\label{fig3}(a) Variation of the fidelity $F(T)$ of mapping the state 
of the atom A in the cavity C$_1$ 
with $\kappa/g$. We have assumed that the cavity decay rates 
are the same for both the modes and $\Delta=10g$.
(b) Variation of the fidelity $F$ calculated at time $2T+\tau$, with the time-delay $\tau$ between 
the atoms for $\kappa=0.002g$ and $\Delta=10g$.}
\end{figure*}

More generally, our QST protocol can be written as
\begin{equation}
|i\rangle_A|i'\rangle_B (\gamma|0,1\rangle+\delta|1,0\rangle)_{\mathrm{cav}}
\stackrel{U(\pi)}{\longrightarrow} 
(\gamma|g\rangle-\delta|f\rangle)_A|i\rangle_B|\psi\rangle_{\mathrm{cav}}\;,
\end{equation}
where $U(\pi)=U_A(\pi)U_B(\pi)$, $U_k(\pi)$ ($k\in $ A,B) $[=\exp\{-iH_{\mathrm{eff}}T/\hbar\}]$ 
denotes the $\pi$-pulse operation on the atom $k$, and 
\begin{equation}
|\psi\rangle_{\mathrm{cav}}=\alpha'|0,1\rangle_{\mathrm{cav}}-\beta'|1,0\rangle_{\mathrm{cav}}\;.
\label{psicav} 
\end{equation}
Our protocol has interesting features: (a) the initial states of the atoms can be
arbitrary, (b) the field state can also be an arbitrary superposition of $|0,1\rangle$
and $|1,0\rangle$.
Note that in case of two-level atom interacting with a resonant single-mode 
cavity, the QST protocol from one atom to another atom has difficulties associated
with relative phase
which can be changed 
either by using a conditional phase shift which is essentially a two-qubit 
operation (see Eq.~(3.8), Ref.~\cite{haroche}) or by applying a resonant microwave
field to the atomic qubit.

We note that if the initial state of the atom B is $|g\rangle$ (or $|f\rangle$)
and the cavity is initially in state $|0,1\rangle$ (or $|1,0\rangle$), then
we can not only transfer the state of atom A to B, but also can interchange 
the states between them.
However, the QST protocol described here 
cannot be interpreted as a SWAP gate, as in usual version of a quantum gate, the atoms A and B 
must interact with the field simultaneously.
We also note that, in the process of coherence transfer between two atoms 
using, for example, the scheme of Ref.~\cite{pelli}, the atoms must be addressed by the pulses simultaneously 
which is basically a local interaction. In the present protocol, the atoms 
interact with the $\pi$-pulse in a sequential manner. This is essentially a 
non-local process.

Extending the idea of QST described above to a number of atoms, we can transfer 
the state of any atom to the consecutive atom. This means, if we consider a 
sequel of atoms, then the state of any atom can be transferred to the consecutive
atom which will pass the cavity after the former leaves the cavity. 
The procedure of transfer of atomic states to consecutive atom 
has been shown schematically in Fig.~\ref{fig2}. 
Here the atoms A, B, C, etc. are sent through another
identical bimodal cavity in initial state $|0,1\rangle$. After passing through 
this cavity, the atom C is again prepared in state $|i\rangle$. Thus, using  
a second cavity in this way, we can transfer the state of the first atom A to a third atom
C. Clearly, if we would use $n$ number of cavities in this sequence, we could 
transfer the state of the atom A to $(n+1)$-th atom in the sequence.

\subsection*{Effects of decoherence - fidelity of QST protocol}
Decoherence is a strong limiting factor in the realization of any quantum computational 
protocol. The interaction of the atom and the cavity with the environment causes
them to decay and results in decoherence. Thus, one has to consider the effect
of decoherence to examine with how much efficiency, the desired outcome can
be produced. These calculations can be done in the density matrix 
framework using the following Liouville equation
\begin{eqnarray}
\dot{\rho}=-\frac{i}{\hbar}[H_{\mathrm{eff}},\rho]&-&\kappa_a(a^\dag a\rho-2a\rho a^\dag +\rho a^\dag a)\nonumber\\
&-&\kappa_b(b^\dag b\rho-2b\rho b^\dag +\rho b^\dag b)\;,
\label{decay}
\end{eqnarray}
where $\kappa_a$ and $\kappa_b$ are the decay constants of the two modes
and $H_{\mathrm{eff}}$ is given by Eq.~(\ref{effham}).

In the present case, to investigate the effect of decoherence, let us consider 
a possible scheme.
We consider $|g\rangle$ and $|f\rangle$ to be the Rydberg levels as in 
Haroche's experiments.
In that case, we 
can use a bimodal microwave cavity like the one used by the group of Haroche.
We use parameters similar to those 
in the experiments by Haroche and his co-workers. If the cavity coupling 
constant $g$ is 
$2\pi\times 50$ kHz and the cavity decay constant $\kappa_a=\kappa_b=\kappa$ 
for each mode is $2\pi\times 100$ Hz,
then $\kappa/g=0.002$. Further, for $\Delta=10g$, we calculate the cavity 
interaction time to be 50 $\mu$s for a $\pi$ pulse, which is consistent with the interaction time 
possible to achieve in a microwave experiment. One sends the atoms with a 
velocity $\sim 10^2$ cm s$^{-1}$ through a few cm long cavity to achieve this 
interaction time. 
Using these parameters, we calculate the fidelity $F$ that the first 
step of the evolution (\ref{qst}) occurs. The variation of $F(T)$
with the decay constant $\kappa$ is shown in Fig.~\ref{fig3}(a),
where $T$ is the interaction time of the atom with the cavity.
Note that the probability that the state of the atom A is transferred to the 
cavity
remains more than $90\%$ for $\kappa=0.002g$. We next show [see Fig.~\ref{fig3}(b)] the variation of the
fidelity $F(2T+\tau)$ of the entire process (\ref{qst}) to occur with the 
time-delay $\tau$ between the atoms A and B for $\kappa =0.002g$.
It is clear that  the probability that the atom B 
acquires the desired state remains above $80\%$ even at 
$g\tau=20 (\equiv\tau\approx 63 \mu$s). 

\begin{figure}
\scalebox{0.32}{\includegraphics{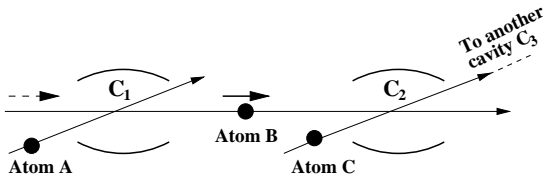}}
\caption{\label{fig4} Schematic diagram for the quantum network between 
distant cavities via atomic channel. Description of the figure is in the text.}
\end{figure}

\section{Extensions of quantum state transfer protocol} 
\subsection{Quantum networks}
Now we show how the above QST protocol can be made useful in preparing  a
quantum network, in which long-lived atomic states are used to communicate between 
the two nodes of the network. We assume that there are two identical two-mode
cavities C$_1$ and C$_2$, which are considered as two nodes of the 
network. Let us consider that the cavity C$_1$ is initially in a state $|0,1\rangle$.
To prepare this cavity in a superposition state 
\begin{equation}
\label{ano}|E\rangle_{\mathrm{cav}}=\alpha |0,1\rangle_{\mathrm{cav}} -\beta |1,0\rangle_{\mathrm{cav}}\;,
\end{equation}
 we send an atom A in 
state $|i\rangle$ through the cavity (see Fig.~\ref{fig4}) such that the atom A
experiences a $\pi$ pulse.
Now our goal is to transfer this cavity state $|E\rangle_{\mathrm{cav}}$ to the other node
C$_2$.
For that we send a second atom B through the cavity C$_1$ after A comes out of it. We
see that the atom B is prepared in $|i\rangle$ state through the evolution (\ref{qst}). This atom is now sent through the second node 
 C$_2$ which is initially in state $|0,1\rangle$. 
In this way, the state $|E\rangle_{\mathrm{cav}}$ of node C$_1$ is 
transferred to the node C$_2$. 

Extending the above idea to a number of distant nodes (cavities), we thus can transfer the state $|E\rangle_{\mathrm{cav}}$ 
from one node to another node of the proposed quantum network via a 
quantum channel (atom). For example, to send this state $|E\rangle_{\mathrm{cav}}$ from 
C$_2$ to another node (say, C$_3$), we can send a third atom C through these two 
nodes subsequently.

We emphasize that our protocol of quantum networking is distinct from 
 the teleportation protocol  of Davidovich {\it et al.\/} \cite{davi}. 
Their protocol depends on the Bell state measurements whereas in our protocol 
no Bell measurement is ever made.

We further note that the present scheme can be used to spread entanglement between two distant cavities. For this, one first sends an atom A in $|g\rangle$ state
through the first cavity C$_1$ prepared initially in the state $|1,0\rangle$ such that the atom experiences a $\pi/2$ pulse ($2g^2T/\Delta=\pi/2$). 
This would prepare the atom and the cavity in the following entangled state:
\begin{equation}
|\Psi\rangle_{AC_1}=\frac{1}{\sqrt{2}}e^{i\pi/2}(|g\rangle_A|1,0\rangle_1+|f\rangle_A|0,1\rangle_1)\;,
\end{equation}
Next the atom passes through 
a second cavity C$_2$ initially in the state $|0,1\rangle$ and experiences a 
$\pi$ pulse. Thus, at the end of this process, the two 
cavities are prepared in an {\it entangled state of two modes\/} as
\begin{equation}
|\Psi\rangle_{C_1C_2}=\frac{1}{\sqrt{2}}e^{i\pi/2}[|1,0\rangle_1|0,1\rangle_2-|0,1\rangle_1|1,0\rangle_2]\;.
\end{equation} 
Clearly one can spread entanglement between atom and the cavity to another distant cavity. Note that in our proposal, 
entanglement is created between the modes of the two different cavities. The entanglement
between two-modes of a single cavity has been produced in \cite{raus}. 

\subsection{Storage and retrieval of an arbitrary superposition state of two modes of a cavity}
We now discuss how the present $\pi$-pulse technique  
can be used to prepare an efficient quantum memory for arbitrary superposition 
of two cavity modes, where there is only one photon is present in either mode. 
Let us consider a two-mode cavity which is in a superposition state of two modes [see Eq.~(\ref{ano})]
\begin{equation}
|E\rangle_{\mathrm{cav}}=\alpha|0,1\rangle_{\mathrm{cav}} -\beta|1,0\rangle_{\mathrm{cav}}\;,
\label{ecav}
\end{equation}
where $\alpha$ and $\beta$ are {\it known\/} coefficients. 
Now we send an 
atom in state (\ref{state2}) through the cavity. 
Applying a $\pi$-pulse on it, we can map the superposition of $|E\rangle_{\mathrm{cav}}$ into the state of the 
atom. This procedure can be written as 
\begin{equation}
|i'\rangle|E\rangle_{\mathrm{cav}}\longrightarrow -|i\rangle|\psi\rangle_{\mathrm{cav}}\;,
\end{equation} 
where $|i\rangle=\alpha|g\rangle+\beta|f\rangle$ and $|\psi\rangle_{\mathrm{cav}}$
is given by Eq.~(\ref{psicav}).
Because, the states $|g\rangle$ and $|f\rangle$ of the atom are radiatively 
long lived, information about the state of the cavity can be stored 
inside the atom for sufficiently long time. To retrieve this 
information into the cavity, 
we prepare a {\it second\/}  cavity in either of the states $|0,1\rangle$ or $|1,0\rangle$ 
and send the atom in state $|i\rangle$ through the 
cavity. Upon applying a $\pi$-pulse, the cavity can again be prepared in the
superposition state as 
before. The retrieval of superposition can be shown as 
\begin{equation}|i\rangle |0,1\rangle_{\mathrm{cav}}\rightarrow |g\rangle |E\rangle_{\mathrm{cav}}\;,\; |i\rangle |1,0\rangle_{\mathrm{cav}}\rightarrow -|f\rangle |E\rangle_{\mathrm{cav}}\;.
\label{decode}
\end{equation}

 
We should mention here that the quantum memory proposed here for the 
cavity state is expected to 
work better since the information is being stored inside the long-lived 
atomic states $|g\rangle$ and $|f\rangle$. However, the transfer time
of the cavity state to the atom is limited by the cavity holding time 
and the atom must stop interacting with the cavity before it decays. We also note that
if the two modes are degenerate and correspond to two states of circular
polarizations, then (\ref{ecav}) can be viewed as a superposition of two
polarization states of a photon. In such a case our proposal corresponds to
storage and retrieval of the polarization states of a photon.

\section{Conclusion}
In conclusion, we have presented a protocol for the transfer of a quantum state
from one atom to another atom. This protocol can be extended to a number of 
atoms passing through sequential cavities and thus one can set up a 
quantum network. We have further shown how an efficient quantum memory 
of arbitrary superposition of two cavity modes can be built up. Our proposals have 
certain advantages as we work with long-lived states of atoms. 
We provide a proper estimate of the efficiency of the state transfer protocol 
against cavity decoherence.

\end{document}